\documentclass[aps, reprint,prl,nofootinbib,longbibliography]{revtex4-1}
\usepackage{dcolumn}
\usepackage[utf8]{inputenc}
\usepackage{amsmath}
\usepackage{amsfonts}
\usepackage{amssymb}
\usepackage{graphicx}
\usepackage{hyperref}
\usepackage{xcolor}

\newcommand{\ket}[1]{\ensuremath{\left|{#1}\right\rangle}}

\newcommand{\mb}{\mathbf}

\begin{document}

\title{Quantum Interference of Force with Entangled Photons}

\author{Gabriela S. Militani}\author{Artur Matoso}\author{Denise F. Ávila}\author{Raul Corrêa} \author{Reinaldo O. Vianna}
\author{Pablo L. Saldanha}\email{saldanha@fisica.ufmg.br}\author{Sebastião Pádua}\email{spadua@fisica.ufmg.br}\affiliation{Departamento de F\'isica, Universidade Federal de Minas Gerais, Belo Horizonte, MG 31270-901, Brazil}
 %\altaffiliation[Also at ]{Physics Department, XYZ University.}%Lines break automatically or can be forced with 

\date{\today}% It is always \today, today,
             %  but any date may be explicitly specified

\begin{abstract}
In this work we experimentally demonstrate the quantum interference of force effect using pairs of entangled photons. Although photons are massless particles, they have linear momentum, and our experiments show that the quantum superposition of a positive momentum transfer with a null momentum transfer may result in a negative momentum transfer to an ensemble of quantum particles (photons), a behavior with no classical analogue. The momentum transfer to each photon is defined by the result of a polarization measurement performed in a second  photon, initially entangled with it. 
\end{abstract}

%\pacs{}

\maketitle

The weak value concept \cite{aharonov88,dressel14} has generated the proposition of many quantum paradoxes in the literature \cite{aharonov88,aharonov13,aharonov13b,vaidman13,aharonov16}. Some of them arrive at controversial conclusions, such as that a neutron could be separated from its spin \cite{aharonov13,denkmayr14}, a photon could propagate in an interferometer through parts where it could not be \cite{vaidman13,danan13}, or that it is possible to put three particles in two boxes without any two particles being in the same box \cite{aharonov16,waegell17,chen19,reznik20}. It is important to stress that such dramatic conclusions are based on a realistic view of the weak values. All the experimental predictions and observations from these works can be explained as quantum interference effects \cite{correa15,atherton15,saldanha14,bartkiewicz15,englert17,correa21}, without such paradoxical behaviors, if we deny a realistic view of the weak values \cite{aredes24}. But some impressive and non-intuitive behaviors of quantum systems that do not depend on a particular interpretation of quantum mechanics were also predicted using the weak values concept. For instance, in Ref. \cite{aharonov13b} Aharonov \textit{et al.} discuss the  radiation pressure that photons exert on mirrors in an interferometer using the weak value concept, showing very interesting results. One conclusion is that the superposition of a positive photonic radiation pressure with a null radiation pressure may result in a negative photonic radiation pressure in a mirror, an extremely non-intuitive behavior.

Inspired in the results of Ref. \cite{aharonov13b}, the quantum interference of force effect was introduced in Ref. \cite{correa18}. It was shown that the quantum superposition of a positive force with a null force on a quantum particle may result in a negative momentum transfer to it. As a consequence, an ensemble of quantum particles may receive an average momentum in the opposite direction of the applied force, a behavior with no classical analogue \cite{correa18}. Due to this effect, it was also shown how charges of the same sign may suffer an effective electrostatic attraction  in an interferometer \cite{cenni19}, another non-intuitive behavior. But these intriguing effects were not experimentally verified so far.

Here we implement a quantum interference of force experiment using pairs of entangled photons generated by spontaneous parametric down- conversion. Photons have linear momentum, even though they have null mass. We experimentally show that the quantum superposition of a positive momentum transfer with a null momentum transfer may result in a negative momentum transfer to an ensemble of photons. The manipulation of the photon momentum is performed with a phase-only spatial light modulator (SLM), which can change the photon propagation direction (and, consequently, its momentum) in a polarization-dependent way. The momentum transfer to each photon is defined by the result of a polarization measurement performed in a second  photon, in a different spatial location, initially entangled with the former. Interestingly, the polarization measurement of the second photon and the consequent definition of the momentum transferred to the first one can be made after the first photon was already detected, in a ``delayed-choice'' scheme.

\begin{figure}
    \centering
    \includegraphics[width=8.5cm]{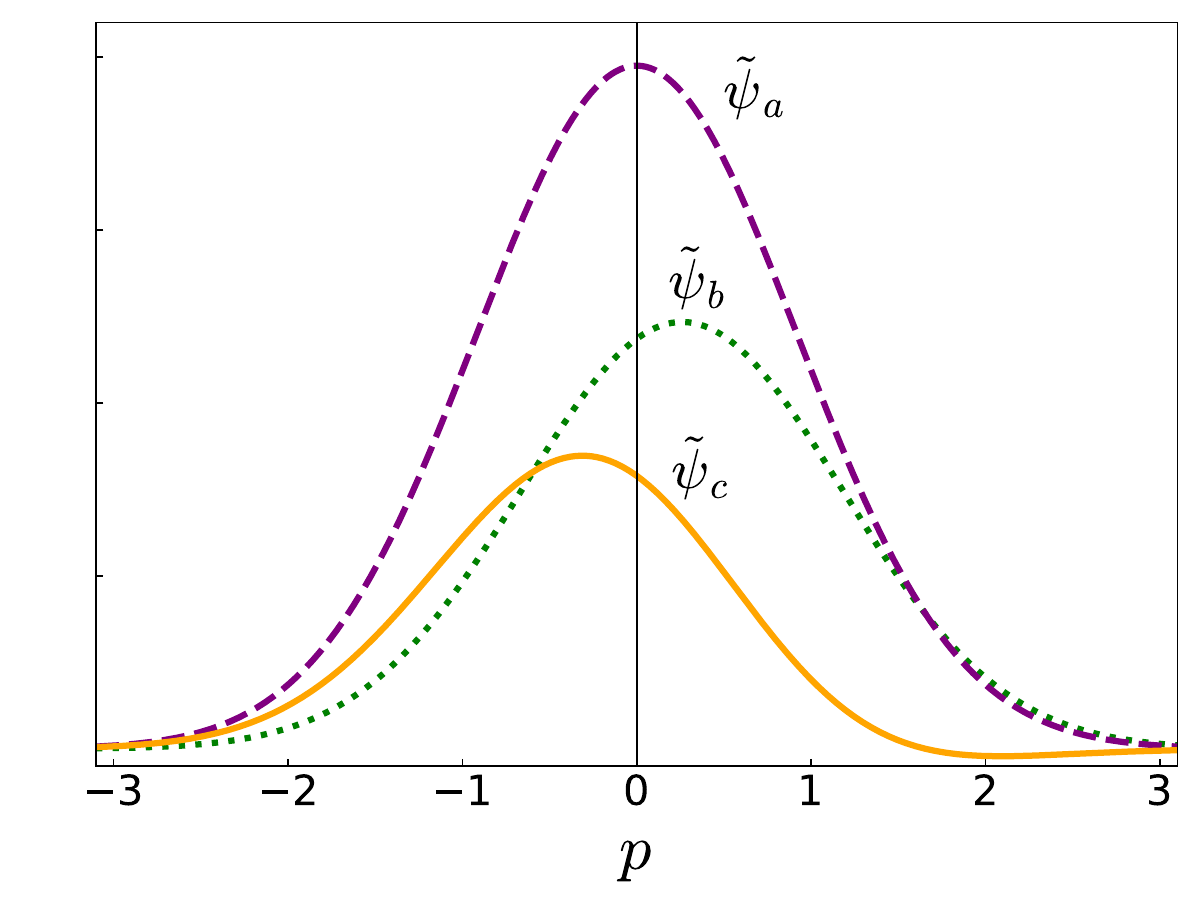}
    \caption{Illustration of a quantum interference of force. The purple dashed line corresponds to the component $\tilde{\psi}_a(p)$ of the momentum wave function of a quantum particle in path $a$ of an interferometer, centered in zero. The green dotted line represents the momentum wave function $\tilde{\psi}_b(p)$ in path $b$, which suffers a force and acquires a positive average momentum.  We assume that both wave functions $\tilde{\psi}_a(p)$ and $\tilde{\psi}_b(p)$ are real. The yellow continuous line corresponds to a wave function $\tilde{\psi}_c(p)=\tilde{\psi}_a(p)-\tilde{\psi}_b(p)$ in one of the interferometer exits, with a negative average momentum.}
    \label{fig:amp}
 \end{figure}

The quantum interference of force is a simple quantum interference effect \cite{correa18}. Consider that a quantum particle whose wave function has a null average momentum in the $x$ direction is sent to an interferometer with two possible paths. In one of the paths, $a$, there is a null force on it, while in the second path, $b$, the particle receives a positive momentum $\delta$ in the $x$ direction smaller than its initial momentum uncertainty. If the probability amplitude that the particle propagates through path $a$ is higher than the one for propagation in path $b$ and we have (partially) destructive interference in one of the interferometer exits, we may have an average momentum transfer in the opposite direction of $\delta$ for particles that leave the interferometer by this exit, as shown in Fig. \ref{fig:amp}. In this figure, $\tilde{\psi}_a(p)$ represents the wave function for the $x$ component of the particle momentum in path $a$, while $\tilde{\psi}_b(p)$ represents the  $x$ component of the particle momentum wavefunction in path $b$, displaced by an amount $\delta$. We assume that both wave functions $\tilde{\psi}_a(p)$ and $\tilde{\psi}_b(p)$ are real. A post-selection may result in a (non-normalized) momentum wavefunction $\tilde{\psi}_c(p)=\tilde{\psi}_a(p)-\tilde{\psi}_b(p)$ %($|\tilde{\psi_c}|^2=1$)
with negative average momentum, as shown in Fig. \ref{fig:amp}. In this situation, the positive momentum components of the wavefunction $\tilde{\psi}_a$ are more subtracted by the wavefunction $\tilde{\psi}_b$ than the negative ones, resulting in this anomalous behavior. Note that this anomalous momentum transfer occurs even outside the weak interaction regime, with $\delta$ being of the same order of magnitude as the particle momentum uncertainty, as in Fig. \ref{fig:amp}. It is important to stress that, by also considering the other exit port of the interferometer, where the particle wavefunction is $\tilde{\psi}_d(p)=\tilde{\psi}_a(p)+\tilde{\psi}_b(p)$, the average momentum transfer is the classically expected one \cite{correa18}. However, in a classical particles system, it would be impossible to post-select an ensemble of particles in a momentum-independent way and obtain an average momentum transfer in the opposite direction of the applied force, as a quantum system permits \cite{correa18}.

\begin{figure}
    \centering
    \includegraphics[width=8.5cm]{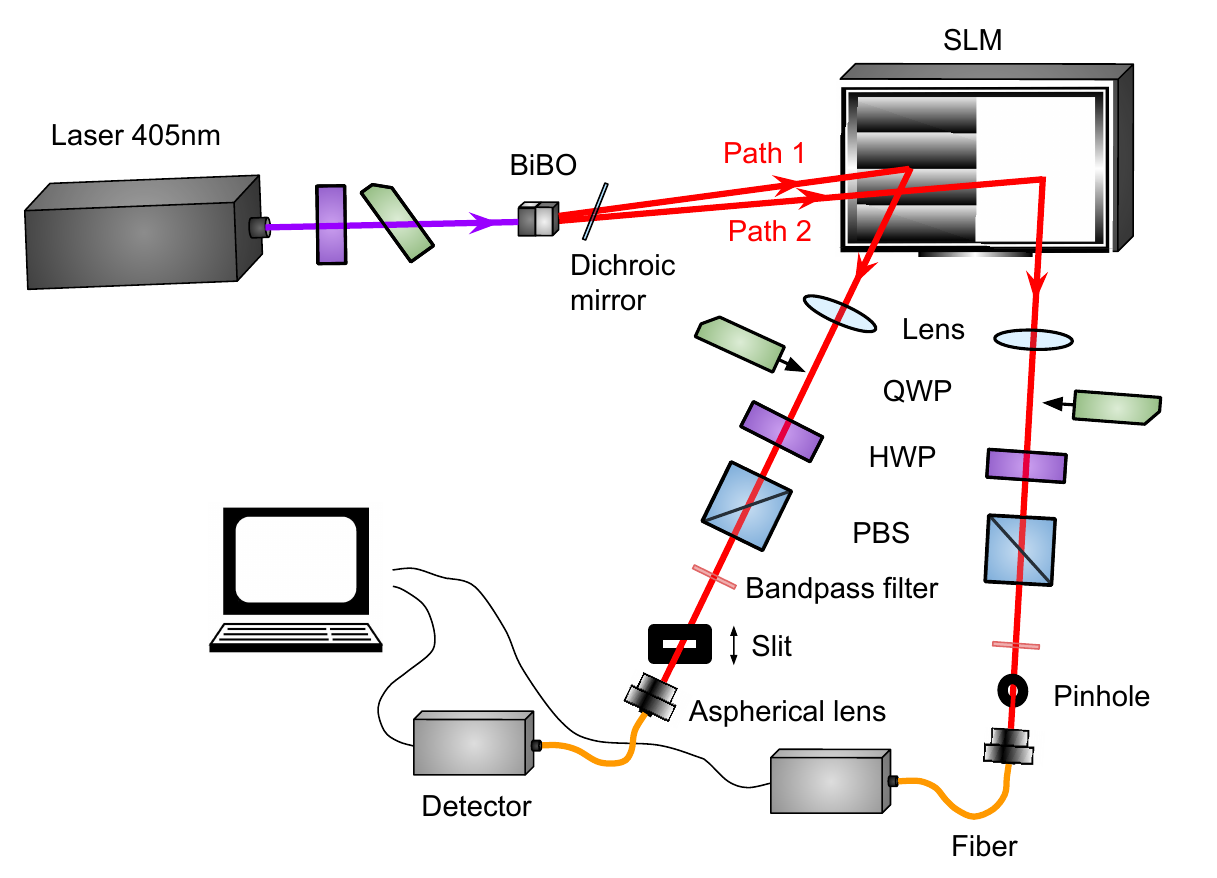}
    \caption{Experimental setup. Entangled photon pairs are generated via spontaneous parametric down-conversion using two BiBO crystals and travel through different paths (1 and 2). A dichroic mirror reflects the incident pump beam, transmitting the down-converted photons. Each photon of the pair is reflected by a different section of the spatial light modulator (SLM) screen, suffering different polarization- and position-dependent phase changes. A slit (path 1) and a pinhole (path 2) are positioned at the focal planes of lenses in the photons paths.  Half-wave plates (HWP) and polarizing beam splitters (PBS) are used to perform polarization projections. Quarter-wave plates (QWP) are inserted in the process of polarization tomography. Bandpass filters select photons with half the frequency of the pump beam. Aspherical lenses couple the photons on each path to an optical fiber to be later on sent to the detection system. This system consists of two single photon detectors and the electronics to register single and coincidence counts and send the data to a computer.}
    \label{fig:setup}
 \end{figure}

Fig. \ref{fig:setup} depicts our experimental scheme. A laser beam with wavelength at  $405$ nm passes through a monomode optical fiber and a polarizing beam-splitter (PBS) to filter the laser beam to a Gaussian profile (not shown in the figure). This pump beam propagates through a half-wave plate (HWP) and a tilted quarter-wave plate (QWP) before reaching a pair of adjacent BiBO crystals with orthogonal optical axes, responsible for generating entangled photon pairs by spontaneous parametric down-conversion. The vertical (V) polarization component of the pump generates pairs of photons with horizontal (H) polarization in one of the crystals, while the pump horizontal component generates photon pairs with vertical polarization in the other \cite{kwiat99}. 
Entanglement arises when it is impossible to distinguish among the crystals which one was responsible for generating the photons. This is the case for the Bell state $\ket{\Phi^+} = [\ket{HH}+\ket{VV}]/\sqrt{2}$ that we sought to achieve, $\ket{HH}$ being a state where both photons have horizontal polarization, while $\ket{VV}$ indicates they have vertical polarization.
For thin crystals, the momentum state of the generated photons is entangled due to the transfer of the angular spectrum of the pump laser to the state of the pair of photons, with the transverse linear momentum being conserved in the process \cite{monken98,walborn10,saldanha13}. The quantum state of the generated photons, considering the polarization and 
%the $\hat{p}_i$ 
one of the components of the transversal momentum, can be written as
\begin{equation}\label{Psi}
    \ket{\Psi}=\int dp_1\int dp_2 \, \tilde{\psi}(p_1+p_2)\ket{p_1,p_2}\otimes\frac{\ket{VV}+ e^{i\phi}\ket{HH}}{\sqrt{2}},
\end{equation}
where $\tilde{\psi}(p)$ describes the pump angular spectrum in a spatial direction transversal to the propagating one, written in terms of the photon momentum \cite{monken98,walborn10,saldanha13}. $\ket{p_{1},p_{2}}$ is a state where $p_1$ corresponds to the momentum of the photon 1 and $p_2$ to the momentum of the photon 2. $\phi$ is a phase that depends of the phase-matching and crystal thickness \cite{kwiat99}. %

The twin photons are reflected by a phase-only spatial light modulator (SLM), that subjects the horizontal polarization component of photon 1 (in path 1) to a linear phase grating, which changes its propagation direction and causes a positive shift $\delta$ on its transverse momentum. At the same time, it does not affect the vertical polarization component, such that  photon 1 suffers a superposition of a positive momentum transfer with a null momentum transfer.
The SLM also subjects the horizontal component of photon 2 (in path 2) to a uniform phase mask, which is set to modify the original phase $\phi$ present in the state of Eq. (\ref{Psi}) to $\pi$. The photon pair quantum state after the SLM then becomes
\begin{equation}
\label{Psislm}
\begin{split}
    \ket{\Psi}= \int dp_1\int dp_2 & \frac{\ket{p_1,p_2}}{\sqrt{2}} 
    \otimes\,\bigl[ \tilde{\psi}(p_1+p_2)\ket{VV}+\\&  - \tilde{\psi}(p_1 -\delta +p_2 )\ket{HH} \bigl].
    \end{split}
\end{equation}

Half-wave plates and polarizing beam splitters are positioned in the paths of both photons to project their polarization states, as shown in Fig. \ref{fig:setup}. Bandpass filters select wavelengths around 810 nm for each photon. In path 1, a slit with $200$ $\mu$m width is mounted at the focal plane of the lens with a focal distance of $40$ cm. In path 2, a pinhole with a diameter of $400$ $\mu$m is positioned at the focus of the lens with a focal distance of $50$ cm. After the slit or pinhole, aspherical lenses are used to fiber couple each photon of the pair to a single photon detector.

At the focal plane of a lens in the paraxial regime, photons transverse momenta are mapped into distances from the propagation axis \cite{saleh}. 
A photon detection at the focal plane of a lens at a distance $x$ from the propagation axis corresponds to the projection of the photon transverse momentum in a state with the $x$ component of the wave vector equal to $k={2\pi}x/({\lambda f})$, $\lambda$ being the photon wavelength and $f$ the lens focal distance \cite{saleh}. Since a photon with wave vector $\mb{k}$ has a momentum $\mb{p}=\hbar\mb{k}$, by detecting photon 1 in the $x$ position using the slit, we have a measurement of the photon 1 $x$-momentum component with the result $p={2\pi\hbar}x/({\lambda f})$. Therefore, by scanning the slit in the path of photon 1  and measuring the photon detection coincidences, we obtain the photon 1 $x$-momentum distribution conditioned to a null transverse momentum for photon 2, selected by the pinhole in path 2.

To simplify our discussion, let us consider that polarization and momentum measurements are performed on photon 2 to prepare the state of photon 1. The order of measurements does not affect the final results. Later we discuss other possibilities, as a delayed-choice scheme where the measurements on photon 1 are performed first. Considering that photon 2 is projected on the state $\ket{p_2=0}\otimes[\sin(\theta_2)\ket{V}+\cos(\theta_2)\ket{H}]$,  by using Eq. (\ref{Psi}) we arrive at the following (non-normalized) quantum state for photon 1:
\begin{equation}\label{Psi1}
    \begin{split}
        \ket{\Psi_1}= \int dp_1 \ket{p_1}\otimes & \bigl[ \sin{(\theta_2)} \tilde{\psi}(p_1 )   \ket{V} +\\
        & - \cos{(\theta_2)} \tilde{\psi}(p_1-\delta)  \ket{H} \bigl].
    \end{split}
\end{equation}
 If the polarization of photon 1 is projected on the state $\sin(\theta_1)\ket{V}+\cos(\theta_1)\ket{H}$ and the angular spectrum of the pump beam is a Gaussian function with $\tilde{\psi}(p_1)\propto\mathrm{Exp}[-p_1^2/(2\sigma^2)]$, the (non-normalized) momentum wave function of photon 1 becomes
\begin{equation}\label{wf}
\tilde{\Psi}_1(p_1)={\sin(\theta_1)\sin(\theta_2)}e^{-\frac{p_1^2}{2\sigma^2}}-{\cos(\theta_1)\cos(\theta_2)}e^{-\frac{(p_1-\delta)^2}{2\sigma^2}}.
\end{equation}
We thus see that the momentum state of photon 1 depends on the polarization measurement performed on photon 2, being the superposition of a momentum wavefunction with null average momentum with a momentum wavefunction with a positive average momentum $\delta$.

The number of coincidence counts $C$ in a time interval for the detectors in the scheme of Fig. \ref{fig:setup} when the position of the slit in path 1 selects a momentum around $p_1$ for photon 1 is proportional to $|\tilde{\Psi}_1(p_1)|^2$. We then consider the following expression to model our experimental results: 
\begin{eqnarray}\nonumber\label{coinc}
C(\theta_1,\theta_2,p_1)=&& A \;[\; \sin(\theta_1)\sin(\theta_2)e^{-\frac{p_1^2}{2\sigma^2}}+\\&&-{\cos(\theta_1)\cos(\theta_2)}e^{-\frac{(p_1-\delta)^2}{2\sigma^2}}\;]^2,
\end{eqnarray}
$A$ being a free parameter related to the photon pair production rate and detection efficiency.

Figure 3 shows our main experimental results. The polarization of photon 1 is always selected in the state $\sin(\theta_1)\ket{V}+\cos(\theta_1)\ket{H}$ with $\theta_1=62^o$.  The green circles correspond to the coincidence counts when the polarization of photon 2 is selected in the vertical direction, in the state $\sin(\theta_2)\ket{V}+\cos(\theta_2)\ket{H}$ with $\theta_2=90^o$. In this case, only the Gaussian with momentum centered in zero is selected on the wavefunction of Eq. (\ref{wf}). 
%A Gaussian fit based on Eq. (\ref{coinc}) using the software OriginPro\texttrademark $\;$gives us the parameters $A=27.3\pm0.7$, $\sigma=5.2\pm 0.1 \; \mu$m$/\hbar$. The origin $p_1=0$ for the horizontal axis was chosen such that the fitted Gaussian is centered on zero. 
The black squares correspond to the coincidence counts when the polarization of photon 2 is selected in the horizontal direction, with $\theta_2=0^o$. In this case, only the Gaussian with momentum centered on $\delta$ is selected on the wavefunction of Eq. (\ref{wf}). The blue triangles represent the coincidence counts when the polarization of photon 2 is selected in the diagonal direction, with $\theta_2=45^o$, where an anomalous momentum transfer can be seen. The average momentum of the detected photons is clearly negative in this case. The plotted curves in Fig. 3 correspond to the predictions of Eq. (\ref{coinc}) with the optimized parameters $A=401$, $\sigma=4.79\hbar/\mathrm{mm}$, and $\delta=2.88\hbar/\mathrm{mm}$, showing a very good agreement with the experimental data.

\begin{figure}
    \centering
    \includegraphics[width=8.5cm]{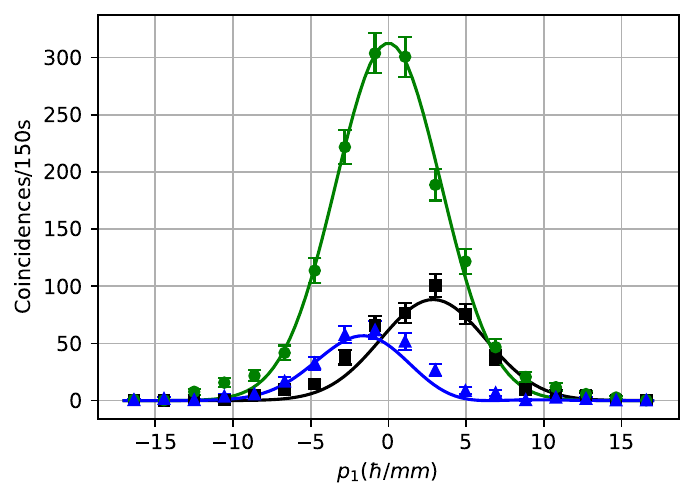}
    \caption{Coincidence counts in the scheme of Fig. 2 when the slit in the path of photon 1 is moved, with each position $x_1$ of the slit being associated to a $x$-component of momentum given by $p_1={2\pi\hbar}x_1/({\lambda f})$, where $\lambda$ is the photon wavelength and $f$ the focal distance of the lens in its path. The polarization of photon 1 is selected in the state $\sin(\theta_1)\ket{V}+\cos(\theta_1)\ket{H}$ with $\theta_1=62^o$ in all measurements. The green circles show the coincidence counts (associated to the photon 1 momentum distribution) when photon 2 is measured with zero momentum and vertical polarization, such that we retrodictively conclude that photon 1 also had an initial vertical polarization and the SLM does not change its momentum. The black squares show the coincidence counts when photon 2 is measured with zero momentum and horizontal polarization, such that we retrodictively conclude that the initial polarization of photon 1 was also horizontal and the SLM transfers a positive momentum $\delta$ to it. The blue triangles show the coincidence counts when photon 2 is measured with zero momentum and a diagonal polarization, such that we retrodictively conclude that photon 1 also had an initial diagonal polarization and the SLM performs a superposition of a positive momentum transfer with a null momentum transfer to it. An anomalous negative average momentum can be seen in this case. The theoretical plots are obtained from Eq. (\ref{coinc}) with the optimized parameters $A=401$, $\sigma=4.79\hbar/\mathrm{mm}$, and $\delta=2.88\hbar/\mathrm{mm}$.}
    \label{fig:Ng2}
 \end{figure}

Its worth mentioning that we performed a tomographic characterization of the initial polarization-entangled state $\rho$ of the photon pair, obtaining a purity Tr$(\rho^2)=0.91$. But the consideration of a pure state, with the predictions of Eq. (\ref{coinc}), already provided a good agreement with the experimental data, as seen in Fig. 3.

We thus see that the superposition of a zero momentum transfer to an ensemble of photons (polarization component $V$) with a positive momentum transfer to this ensemble (polarization component $H$) results in a negative momentum transfer to this ensemble of quantum particles. There is no analogous behavior for such anomalous momentum transfer to a set of classical particles.

The probability of a coincidence detection in the scheme of Fig. \ref{fig:setup} does not depend on the order that the measurements are performed. We can imagine a situation where the momentum measurements on photon 1 are performed by a screen sensible to single photons at the focal plane of the lens in path 1. Imagine that, for each detected photon, its twin is kept for a later measurement. In this situation, after many photons were detected by the screen, we could choose in which polarization basis their partners would be measured, choosing the possible momentum transfers for the already detected photons. Of course, for each polarization basis for photon 2  there are two possible outcomes, corresponding to two different momentum transfers to photon 1. After this procedure, we could separate the already detected photons in two ensembles, each one that suffered a different momentum transfer, with one of them being anomalous. Our experimental scheme is actually closer to this ``delayed-choice'' one, since photon 2 is detected slightly after photon 1. In this sense, we could say that the anomalous momentum transfer to photon 1 is determined after it was already detected, by the measurement performed on photon 2.

To conclude, we have implemented a quantum interference of force experiment with photons. One of the photons of an entangled pair suffers a quantum superposition of null and positive momentum transfers. The resultant momentum transfer was shown to have a negative anomalous value, depending on the result of a polarization measurement performed in the other, distant, photon, showing a non-intuitive quantum behavior of Nature, impossible to be implemented with classical particles.

\begin{acknowledgments}
    \textit{Acknowledgments} - Funding was provided by Conselho Nacional de Desenvolvimento Científico e Tecnológico (CNPq, grant 22300/2021-7), Coordenação de Aperfeiçoamento de Pessoal de Nível Superior (CAPES),  Instituto Nacional de Ciência e Tecnologia de Informação Quântica (INCT-IQ 465469/2014-0) and Fundação de Amparo à Pesquisa do Estado de Minas Gerais (FAPEMIG, grant 02718-24 and grant BPD-00996-22).
\end{acknowledgments}

%\bibliography{biblio}% Produces the bibliography via BibTeX.

%

%merlin.mbs apsrev4-1.bst 2010-07-25 4.21a (PWD, AO, DPC) hacked
%Control: key (0)
%Control: author (8) initials jnrlst
%Control: editor formatted (1) identically to author
%Control: production of article title (-1) disabled
%Control: page (0) single
%Control: year (1) truncated
%Control: production of eprint (0) enabled

%Entanglement appears due to the fundamental indistinguishability about in which crystal the photon pair was generated, and a maximally entangled polarization state is produced with the pump beam having equal horizontal and vertical components, which is the situation we use in our experiments.

\end{document}